\documentclass[preprint,aps]{revtex4}

\begin{document}

\title{ Origin of the Immirzi Parameter}

\author{Chung-Hsien Chou}\email{chouch@phys.sinica.edu.tw}
 \affiliation{Institute of Physics, Academia Sinica, Taipei 115, Taiwan}
\author{Roh-Suan Tung}\email{tung@shnu.edu.cn}
  \affiliation{Center for Astrophysics, Shanghai Normal University,
 Shanghai 200234, China}
\author{Hoi-Lai Yu}\email{hlyu@phys.sinica.edu.tw}
 \affiliation{Institute of Physics, Academia Sinica, Taipei 115, Taiwan£¬ ROC}


\begin{abstract}
Using quadratic spinor techniques we demonstrate that the Immirzi
parameter can be expressed as ratio between scalar and pseudo-scalar
contributions in the theory and can be interpreted as a measure of
how Einstein gravity differs from a generally constructed covariant
theory for gravity. This interpretation is independent of how
gravity is quantized. One of the important advantage of deriving the
Immirzi parameter using the quadratic spinor techniques is to allow
the introduction of renormalization scale associated with the
Immirzi parameter through the expectation value of the spinor field
upon quantization.
\end{abstract}

\pacs{04.60.Ds}%

\keywords{}

\maketitle

\section{Introduction}

One of the most direct ways of approaching the quantization of
Einstein's theory of gravity is to put it into a Hamiltonian form
and then try to apply the procedures of canonical quantization.
The fact that Einstein's theory is generally covariant makes the
task one of most difficult problems in theoretical physics if not
impossible. The complicated non-polynomial structure found for the
standard Hamiltonian for general relativity raises another
challenge to researchers. However, Ashtekar managed to make
progresses using new canonical variables to reduce the constraints
to polynomial form. The so called Ashtekar connection \cite{A,RT}
$A= \Gamma + i K$ has a part $\Gamma$ refers to intrinsic
curvature on a spacelike 3-surface $S$, and another part $K$ that
refers to the extrinsic curvature to the spacetime $M$. Because of
several technical difficulties, in order to make progress, a more
general Barbero connection\cite{Ba} was introduced, $ A= \Gamma +
\gamma K $ where $\gamma$ is an arbitrary complex number. It is
also known as the Immirzi parameter\cite{Im} (usually assumed to
be a real number for $\it SU(2)$ Barbero connection).

The important achievement of quantizing the Ashtekar-Barbero
connection variable is the construction of a kinematic Hilbert
space using spin networks\cite{RT}. With spin networks, the area
and volume spectra can be derived. If a spin network intersects a
surface $S$ transversely, then this surface has a definite area in
this state, given as a sum over the spins $j$ of the edges poking
through $S$:
$$ \hbox{Area}(S)=8\pi
\gamma \sum_j \sqrt{j(j+1)}
$$
in units where the $\hbar=c=G=1$, with a free Immirzi parameter
$\gamma$. Due to the presence of the Immirzi parameter, the famous
Bekenstein-Hawking entropy formula, $ S_{BH}= {A \over 4} $ could
not be uniquely determined. This has been viewed as the main
unsatisfactory point of this approach for some years.

Recently Dreyer \cite{DR} proposed a way to fix the Immirzi
parameter using asymptotic behavior of the quasinormal modes of a
Schwarzschild black hole. The result fixed the value $ \gamma =
{\ln 3 \over 2 \sqrt{2} \pi} $ with the lowest possible spin
$j_{min}=1$. Domagala, Lewandowski and Meissner\cite{DLM} fixed an
incorrect assumption that only the minimal value of the spin
contributes. Their result involves the logarithm of a
transcendental number instead of the logarithms of integers; $
\gamma = 0.2375329... (>{\ln 2 \over \pi}) $ with
$j_{min}={1\over2}$. The calculation works for charged and
rotating black holes and black holes coupled to a dilaton field,
with the {\it same} value of $\gamma$. There appears to be no
clear geometrical reason for a particular choice of the real
number value $\gamma$ and obscures its physical interpretation.

The appearance of the Immirzi parameter $\gamma$ can be seen in
the simplest tetrad-Palatini action of general relativity where
one can add an additional term with coupling coefficient $\gamma$.
This newly added term does not  affect the equations of motion. In
the case where torsion free connection that solves the equation of
motion is employed to obtain the Einstein-Hilbert gravity action,
the additional term in the action becomes identically zero.
Arguing with this observation in mind , the effect of $\gamma$ is
therefore, not an physical observable in Einstein gravity.
Recently Perez and Rovelli \cite{PR} had argued that one can
observe physical effects of the Immirzi parameter $\gamma$ by
coupling Einstein gravity to fermionic degrees of freedom. The
presence of matter field induces a torsion term in the connection
and the additional term becomes non-vanishing. Freidel, Minic and
Takeuchi \cite{FMT} discussed parity violation and studied the
coupling of fermionic degrees of freedom in the presence of
torsion from the viewpoint of effective field theory. The
importance of these works are to notice that physical effects
arise from the Immirzi parameter $\gamma$ is measurable and
independent from how gravity is quantized. We believe however, if
the Immirizi parameter $\gamma$ is a physical property of the
gravity sector, then it should be observable without the
introduction of other matter field. In the following, we shall
introduce a Quadratic Spinor Representation of General Relativity
\cite{NT,TJ, TN99} formalism where the physical meanings and
effects of the Immirzi parameter $\gamma$ become transparent in
general relativity. In this formalism, the Immirzi parameter
becomes a ratio between scalar and pseudo-scalar contributions in
the theory and measures how a generally formulated general theory
of relativity differs from Einstein gravity. More importantly, one
can acquire this ratio a renormalization scale upon quantization.

\section{Quadratic spinor representation of General Relativity}

The canonical formulation of Loop Quantum Gravity can be derived by
the Holst action{\cite{Holst},
\begin{equation}
S[\vartheta,\omega]=\alpha \int \ast (\vartheta^a \wedge
\vartheta^b) \wedge R_{ab}(\omega) + \beta \int \vartheta^a \wedge
\vartheta^b \wedge R_{ab}(\omega),
\end{equation}
where the Immirzi parameter is,
\begin{equation}
\gamma = { \alpha\over\beta }.
\end{equation}
In the above, $a, b ... = 0, 1, 2, 3$ being the internal indices
of the internal orthonormal frame. The field  $\vartheta^a$ being
the tetrad field; $\omega$ is the $SU(2)$ connection; $R_{ab}$
being the curvature of $\omega$ and  $\ast$$R$ being its dual.
This is comparable with the Quadratic Spinor Lagrangian
\cite{NT,TJ},
\begin{equation}
{\cal L}_{\psi} = 2D(\overline\psi\vartheta)
\gamma_5D(\vartheta\psi),
\end{equation}
where $\vartheta = \vartheta^a \gamma_a$ and $\gamma_a$ being the
Dirac gamma matrices. The auxiliary spinor field $\psi$ in the
Quadratic Spinor Lagrangian was first introduced by Witten
\cite{Witten} as a convenient tool used in the proof of positive
energy theorem in Einstein gravity.  In a more general context,
this auxiliary spinor field provides a nice gauge condition to
pick up the relevant variables in the theory. The key to these
successes is a ``spinor-curvature identity":
\begin{eqnarray}
&&2D(\overline\psi\vartheta) \gamma_5 D(\vartheta\psi)\nonumber\\
&=&\overline\psi\psi R_{ab}\wedge \ast(\vartheta^a\wedge\vartheta^b)
 +\overline\psi\gamma_5 \psi
R _{ab}\wedge \vartheta^a\wedge\vartheta^b \nonumber\\
&&+d[D(\overline\psi\vartheta)\gamma_5 \vartheta\psi+
\overline\psi\vartheta \gamma_5 D(\vartheta\psi)]
\end{eqnarray}
Note  that in the above expression the boundary terms provide an
important condition in obtaining a finite action at spatial
infinity and consequently a well defined Hamiltonian. Here, the
spinor field $\psi$ plays a key rule which allows one to pick up
the correct gauges in obtaining a well defined Hamiltonian of the
theory. The equation of motion for the connection
$\omega[\vartheta]$ is:
\begin{equation}
D[\overline\psi\psi \ast (\vartheta^a
\wedge\vartheta^b)+\overline\psi \gamma_5 \psi (\vartheta^a
\wedge\vartheta^b)]=0,
\end{equation}
where $D$ is the covariant derivative defined by the connection
variable $\omega[\vartheta]$.  For $\overline\psi \psi = 1$ and
$\overline\psi\gamma_5\psi =0$ the torsion free spin connection
$\omega[\vartheta]$ of the tetrad field $\vartheta$ solves the above
field equation. Therefore, if we set
\begin{equation}
\gamma={\overline\psi\psi \over \overline \psi \gamma_5 \psi}
\end{equation}
then the choice of $\gamma= \infty$ ($\overline\psi\psi=1$ and
$\overline\psi\gamma_5\psi=0$), is the Einstein-Hilbert action in
equation(1); and $\gamma = i$ is the self-dual action in the
Ashtekar canonical gravity framework, and $\gamma = 1$ corresponds
to the action for the Hamiltonian considered by Barbero \cite{Ba}.
The Immirzi parameter $\gamma$ in this setting becomes a measure of
how Einstein gravity differs from a most generally formulated
gravitation theory which satisfies general coordinate covariance. It
is also the ratio between scalar and pseudo-scalar contributions in
the theory as can be seen from the explicit expression of $\gamma$.
Another important feature revealing in this derivation is the
possibility of introducing a renormalization scale $\mu$ associated
with the Immirzi parameter $\gamma$ upon quantization where
expectation of $<\overline\psi  \psi>_\mu$ and
$<\overline\psi\gamma_5\psi>_\mu$ at some scale $\mu$ should be
employed. Thus the ``Quadratic Spinor Representation of General
Relativity" provides a transparent interpretation of the Immirzi
parameter.

A technical drawback of the above derivation is that
$\overline\psi \gamma_5 \psi$ is not in general a real function.
This can be easily seen from using a particular representation of
the Dirac algebra. In order for $\overline\psi \gamma_5 \psi$ to
be always real to render a corresponding real Ashtekar-Barbero
variable, one has to use anti-commuting spinor. In the next
section, we shall develop a model which provides a systematic way
of obtaining a Quadratic Spinor Representation with an appropriate
symmetry for the action where anti-commuting spinor arises
naturally.

\section{The origin of Immirzi parameter}

In this section, we derive the Einstein-Hilbert action in a more
systematic way with anti-commuting spinors using $Osp(1,2C)$
algebra, which is the simplest supersymmetric extension of
$SL(2,C)$ (or $su(2)$) algebra. The algebra has bosonic generators
$J_{00},J_{01}=J_{10},J_{11}$ and fermionic generators $Q_0,Q_1$
which satisfies the following algebra:
\begin{eqnarray}
\left[ J_{AB}, J_{CD} \right] &=&
      \epsilon_{C(A} J_{B)D}
     +\epsilon_{D(A} J_{B)C} , \\
\left[ J_{AB}, Q_C \right] &=& \epsilon_{C(A} Q_{B)} ,\\
\{Q_A, Q_B\} &=& J_{AB} .
\end{eqnarray}
Their complex conjugates $J_{A'B'}$ and $Q_{A'}$ satisfy the same
algebra as above. The $SO(1,3)$ generators $J_{ab}$ can be
constructed by
$J_{ab}=\sigma^{AA'}_a\sigma_b^{B}{}_{A'}J_{AB}+\sigma_a^{AA'}\sigma_{bA}{}^{B'}J_{A'B'}$
where $\sigma_a^{AA'}$ are the Pauli matrices\footnote{The
upper-case Latin letters $A,B,...=0,1$ denote two component spinor
indices, which are raised and lowered with the constant symplectic
spinors $\epsilon_{AB}=-\epsilon_{BA}$ together with its inverse
and their conjugates according to the conventions
$\epsilon_{01}=\epsilon^{01}=+1$,
$\lambda^A:=\epsilon^{AB}\lambda_B$,
$\mu_B:=\mu^A\epsilon_{AB}$.}. The $Osp(1,2C)$ algebra has a
nondegenerate Killing form and the Cartan-Killing metric
$\eta_{\alpha\beta}={\rm{diag}}(\eta_{(AB)(MN)},\eta_{AB})$ is
given by
\begin{eqnarray}
\eta_{(AB)(MN)}&=& \textstyle{1\over2}
(\epsilon_{AM}\epsilon_{BN}+\epsilon_{AN}\epsilon_{BM}), \\
\eta_{AB}&=& - \epsilon_{AB} .
\end{eqnarray}

Now, one can follow a well defined procedure to construct a
$Osp(1,2C)$ invariant Lagrangian which is quadratic in the spinor
representation. To each generator $T_\alpha=\{J_{AB}, J_{A'B'},
Q_A, Q_{A'} \}$, we associate a 1-form field
$A^\alpha=\{\omega^{AB},\omega^{A'B'},\varphi^{A},\varphi^{A'}\}$,
and construct a super Lie algebra valued connection 1-form,
\begin{equation}
A=A^\alpha T_\alpha = \omega^{AB} J_{AB} +\varphi^A Q_A +c.c. ,
\label{1}
\end{equation}
where $\omega^{AB}$ is the SL(2,C) connection 1-form and
$\varphi^A$ is an anti-commuting spinor valued 1-form. The
curvature is given by $F=dA + \textstyle{1\over2}[A, A]=
dA+\textstyle{1\over2}A^\alpha\wedge A^\beta \otimes [T_\alpha,
T_\beta]$. Given the $Osp(1,2C)$ connection $A$ defined in
equation (\ref{1}), the curvature ($F=F(J)^{AB} J_{AB}+F(Q)^A Q_A
+c.c.$ ) contains a bosonic part associated with $J_{AB}$  and
$J_{A'B'}$,
\begin{equation}
F(J)^{AB}=d\omega^{AB}+\omega^{AC}\wedge \omega_C{}^B +{1\over2}
\varphi^A\wedge\varphi^B ;
\end{equation}
and a fermionic part associated with $Q_A$ and $Q_{A'}$ ,
\begin{equation}
F(Q)^A=d\varphi^A+\omega^{AB}\wedge \varphi_B .
\end{equation}

The action, quadratic in the curvature, using this $Osp(1,2C)$
connection $A$ is
\begin{eqnarray}
{\cal S}_{\rm{T}}[A^\alpha]&=&\int\, F^\alpha\wedge F^\beta \,\eta_{\alpha\beta} \nonumber\\
&=&\int\, F(J)^{AB}\wedge F(J)_{AB} \nonumber\\
&&\qquad + F(Q)^{A}\wedge F(Q)_{A}  +c.c.,
\end{eqnarray}
where $\eta_{pq}$ is the Cartan-Killing metric of the $Osp(1,2C)$
group. However, this action is a total differential and therefore,
is a pure topological action without local dynamics. Hence,
similar to the work of MacDowell and Mansouri \cite{MM}, we break
the topological field theory of this $Osp(1,2C)$ symmetry into its
bosonic sector and fermionic sector. A way to do this is to choose
$i_{\alpha\beta}={\rm{diag}}(i_{(AB)(MN)},i_{AB})$ such that
\begin{eqnarray}
i_{(AB)(MN)}&=& 0, \\
i_{AB}&=& -\epsilon_{AB} .
\end{eqnarray}
The new action is
\begin{eqnarray}
{\cal S}[A^\alpha]&=&\int\, F^\alpha\wedge F^\beta \, i_{\alpha\beta} \nonumber\\
&=&\int\,  F(Q)^{A} \wedge F(Q)_{A} +c.c. \nonumber\\
&=& \int\, D\varphi^A\wedge D\varphi_A  +c.c. \label{action}
\end{eqnarray}
The field equations can be obtained by varying the Lagrangian with
respect to the gauge potentials---the $Osp(1,2C)$ connection. With
these gauge potentials fixed at the boundary, the field equations
are
\begin{eqnarray}
&& D^2\varphi^A=R^{AB}\wedge\varphi_B=0 ,\label{fe1} \\
&&D(\varphi^A \wedge \varphi^B)=0 .\label{fe2}
\end{eqnarray}
plus their corresponding complex conjugate equations. We look for
classical torsion free Einstein solution $R_{ab}$, by making the
ansatz \cite{BM} such that
\begin{equation}
\varphi^A=\vartheta^{AA'} \xi_{A'}\label{2}
\end{equation}
where $\vartheta^{AA'}$ is the tetrad 1-form field (assuming to be
real) and $\xi_{A'}$ is an arbitrary nonzero anti-commuting spinor
field. With the ansatz, the second field equation (\ref{fe2})
gives
\begin{equation}
D(\vartheta^{AA'} \wedge \vartheta^{B}{}_{A'} \xi^{B'}\xi_{B'}
)=0, \label{fe2a}
\end{equation}
which together with their complicated conjugate part, implies the
connection is torsion free. Thus the field equation (\ref{fe1})
becomes
\begin{equation}
R^{AB}\wedge\vartheta_B{}^{A'} \xi_{A'}=0,\label{fe1a}
\end{equation}
together with their complex conjugate equations
\begin{equation}
R^{A'B'}\wedge\vartheta^{A}{}_{B'} \xi_{A}=0;\label{fe1b}
\end{equation}
we derive the Einstein equation
\begin{equation}
R^{ab}\wedge\vartheta_b=0 ,
\end{equation}
where
$R_{ab}=\sigma^{AA'}_a\sigma_b^{B}{}_{A'}R_{AB}+\sigma_a^{AA'}\sigma_{bA}{}^{B'}R_{A'B'}$
and $\vartheta^a=\sigma^a_{AA'} \vartheta^{AA'}$. For a general
formulation which satisfies equation (\ref{2}), the action reduces
to
\begin{eqnarray}
 {\cal S}&=&\int\,
(R_{AB}\wedge \vartheta^{AA'} \wedge \vartheta^B{}_{A'} \xi^{B'}\xi_{B'})+c.c. +d(...) \nonumber\\
&=&\int\, \overline\psi\psi R^{ab}\wedge \ast (\vartheta_a \wedge
\vartheta_b)  +\overline\psi\gamma_5\psi R^{ab}\wedge (\vartheta_a
\wedge \vartheta_b) \nonumber\\ &&
+d[D(\overline\psi\vartheta)\gamma_5 \vartheta\psi+
\overline\psi\vartheta \gamma_5 D(\vartheta\psi)],
\end{eqnarray}
where $\overline\psi\psi=\xi^A\xi_A+\xi^{A'}\xi_{A'}$,
$\overline\psi\gamma_5\psi=i(-\xi^A\xi_A+\xi^{A'}\xi_{A'})$ and
$d(...)$ are the boundary terms which are essential for the action
being well-defined at spatial infinity.

The Immirzi parameter is then given by
\begin{eqnarray}
\gamma={ <\overline\psi\psi> \over <\overline\psi\gamma_5  \psi>} .
\end{eqnarray}
Since both $<\overline\psi \gamma_5 \psi>$ and $<\overline\psi\psi>$
are real for anti-commuting spinor, $\gamma$ is always real.  Here
we have explicitly used the expectation values of the spinor field.
It highlights the renormalization scale dependent nature of $\gamma$
upon quantization in this setting. Same as in the previous session,
from the explicit expression for $\gamma$, it is the ratio between
scalar and pseudo-scalar contributions in the theory and can be
interpreted as a measure of how a generally formulated covariant
theory differs from Einstein gravity. In a future publication we
shall introduce the dynamics of the spinor field $\psi$ and
investigate how $\gamma$ is connected with other properties of
general relativity.

\section{Concluding remarks}
In summary, we have demonstrated how Quadratic Spinor Representation
of general theory of relativity with auxiliary spinor field $\psi$
can provide a systematic way to derive the physical properties of
the Immirzi parameter $\gamma$. From the explicit expression for
$\gamma$, one can see that it is the ratio between scalar and
pseudo-scalar contributions in the theory. An important feature of
this derivation is the possibility of introducing a renormalization
scale associated with $\gamma$ upon quantization.

\vfill

\begin{acknowledgments}
RST is supported by NSFC under grant numbers 10375081 and 10375087.
RST and HLY wants to thank J. Nester for helpful discussions.
\end{acknowledgments}



\begin{thebibliography}{}


\bibitem{A}
A. Ashtekar,
        {\it Phys. Rev. Lett. \bf 57} 2244 (1986);
       {\it Phys. Rev. D \bf 36} 1587 (1987).


\bibitem{RT}For a reveiw see: C Rovelli, Quantum Gravity {\it(Cambridge University
Press, 2004)}; T. Thiemann, Modern Canonical General Relativity.
{\it(Cambridge University Press, 2005)}.

\bibitem{Ba}
F. Barbero, {\it Phys. Rev. D\bf{51}} 5507 (1995); {\it Phys. Rev.
D\bf{51}} (1995) 5498.
\bibitem{Im}
G. Immirzi, {\it Class. Quantum Grav. \bf 14}, L177  (1997).

\bibitem{DR}O. Dreyer, {\it Phys. Rev Lett. \bf{90}} 081301 (2003); gr-qc/0211076.
\bibitem{DLM}M. Domagala, J. Lewandowski, {\it Class. Quantum
Grav. \bf{21}} (2004) 5233; gr-qc/0407051; K. A. Meissner, {\it
Class. Quantum Grav. \bf{21}} (2004) 5245; gr-qc/0407052.

\bibitem{PR} A. Perez and C. Rovelli, ``Physical effects of the Immirzi parameter'',
gr-qc/0505081.

\bibitem{FMT} L. Freidel, D. Minic and T. Takeuchi, ``Quantum
gravity, torsion, parity violation and all that'', hep-th/0507253.


\bibitem{NT}J. M. Nester and R. S. Tung, {\it Gen. Rel. Grav. \bf 27 }, 115 (1995).

\bibitem{TJ} R. S. Tung and T. Jacobson,
{\it Class. Quantum Grav. \bf 12}, L51  (1995).

\bibitem{TN99}
R. S. Tung and J. M. Nester,
        {\it Phys. Rev. D \bf 60}, 021501 (1999).



\bibitem{Holst} S. Holst, {\it Phys. Rev. D \bf 53}, 5966 (1996).


\bibitem{Witten} E. Witten, {\it Commun. Math. Phys. \bf 80,}, 381(1981).

\bibitem{MM}
S. W. MacDowell and F. Mansouri, {\it ``Unified geometric theory
of gravity and supergravity''}; {\it Phys. Rev. Lett. \bf 38}
(1977) 739; Erratum, {\it ibid.} {\bf 38}  (1977) 1376.
\bibitem{BM}
I. Bars and S. W. MacDowell, {\it Phys. Lett. B  \bf 71}, 111
(1977); {\it Gen. Rel. Grav.  \bf 10}, 205 (1979).

\end{thebibliography}
\end{document}